\documentclass[journal]{IEEEtran}
\IEEEoverridecommandlockouts
\usepackage{cite}
\usepackage{amsmath,amssymb,amsfonts}
\usepackage{algorithm,algorithmic}
\usepackage{bm,bbm}
\usepackage{booktabs}
\usepackage{graphicx}
\usepackage{textcomp}
\usepackage{xcolor}
\usepackage{subfigure}
\usepackage[normalem]{ulem}

\newtheorem{theorem}{Theorem}
\newtheorem{Def}{Definition}
\newtheorem{rem}{Remark}
\def\BibTeX{{\rm B\kern-.05em{\sc i\kern-.025em b}\kern-.08em
    T\kern-.1667em\lower.7ex\hbox{E}\kern-.125emX}}
\begin{document}

\title{Age of Information Optimization in Multi-Channel Network with Sided Information\\}

\author{
	Yuchao~Chen,~Jintao~Wang,~\IEEEmembership{Senior Member,~IEEE}, Xiaoqing Wang, and~Jian~Song,~\IEEEmembership{Fellow,~IEEE}\\
	
	\thanks{Y. Chen, J. Wang, J. Song are with Beijing National Research Center for Information Science and Technology (BNRist) and the Department of Electronic Engineering, Tsinghua University, Beijing 100084, China. J. Wang  and J. Song are also with Research Institute of Tsinghua University in Shenzhen, Shenzhen, 518057. X. Wang is with China Mobile Communications Group. (e-mail: \{cyc20@mails.; wangjintao@; jsong@\}tsinghua.edu.cn, 15801215882@139.com) This work was supported in part by Tsinghua University-China Mobile Research Institute Joint Innovation Center. \emph{(Corresponding author: Jintao Wang)}}
}

\maketitle

\begin{abstract}
	We consider a discrete-time multi-channel network where the destination collects time-sensitive packets from multiple sources with sided channel information. The popular metric, Age of Information (AoI), is applied to measure the data freshness at the destination. Due to the interference constraint, only disjoint source-channel pairs can be chosen for transmission in each time slot, and the decision maker should choose the optimal scheduling pairs to minimize the average AoI at the destination. To learn the optimal channel selection, we apply the linear contextual bandit (LCB) framework by utilizing the sided information provided by pilots. Concretely, we establish the relationship between AoI regret and sub-optimal channel selection times and propose both age-independent and age-dependent algorithms. The former method is proven to achieve the sub-linear AoI regret but is outperformed by the latter algorithm both in the linear and non-linear contextual model in simulation.
\end{abstract}

\begin{IEEEkeywords}
	Age of Information, contextual bandit, online learning
\end{IEEEkeywords}

\section{Introduction}
The information freshness has become significantly important in the state updating systems including the Internet of Things (IoT) and Internet of Vehicles (IoV). In these scenarios, the destination collects the fresh information from multiple sources for decision making and state monitoring. The popular Age of Information (AoI) has been proposed \cite{AoI_intro_2012} to capture this data freshness for the destination. Since then, numerous studies have been conducted to design optimal scheduling algorithms to minimize AoI in the wireless networks.

A large amount of research have studied the scheduling decisions in the single-hop multi-source networks with known statistics \cite{thyJASC2020,cyc2020entropy,TMC2021Eytan}. In \cite{thyJASC2020,cyc2020entropy}, the age minimization problem is converted into a constrained Markov decision process (CMDP) and a linear programming optimization. Further considering random packet arrival and different queueing disciplines, \cite{TMC2021Eytan} proposes a low-complexity Max-Weight policy, which is shown to closely achieve the analytical lower bound.

However, the aforementioned methods require channel statistics such as the successful transmission probability and transition matrix of channel states. In reality, these parameters are hard to obtain in advance due to the time varying channel condition. To overcome the challenge, online learning methods have been incorporated into the original scheduling algorithms to both adaptively learn these parameters and select the optimal source or channel for age optimization \cite{AoIbandit2020,agebandit2021,LiBin_Infocom_2021}. In \cite{AoIbandit2020,agebandit2021}, the authors cast the AoI minimization problem into the multi-armed bandit (MAB) framework. The standard upper-confidence-bound (UCB) and Thompson sampling (TS) methods have been modified to guide the channel learning algorithm. Theoretically these methods guarantee the sub-linear AoI regret compared with the optimal policy in hindsight.

In above learning algorithms, the decision maker only has the historical observation and decisions for scheduling. However, in many communication scenarios, the central controller can obtain some sided channel information such as SNR or SINR from pilots or training symbols \cite{CB2021Jung,Wang2022CB}. To model the sided information, these works apply the contextual bandit framework \cite{Zhou2015CBsurvey}. In \cite{CB2021Jung}, the authors use the linear function to model the reward while in \cite{Wang2022CB}, a multi-player bandit is considered for decentralized learning. However, these algorithms cannot be applied to satisfy the favorable age performance because of the Markovian property of AoI. Therefore, new algorithms should be designed for this general framework.

To our best knowledge, this is the first work to optimize AoI in the multi-source multi-channel network with sided channel information. In each slot, the packet of each source comes in random, and the controller can choose disjoint source-channel pairs for transmission due to the interference constraint. First, we theoretically prove that the AoI regret grows at the same order of accumulative sub-optimal channel selection times under the optimal source selection policy. Then for the channel selection policy, we incorporate the contextual bandit framework into the channel learning algorithm. We propose both age-independent and age-dependent algorithms for channel selection, and show the superiority of considering the current AoI into the decision making.

\section{Problem Formulation}
\subsection{System Model}
\begin{figure}[ht]
	\centering
	\includegraphics[width=.8\columnwidth]  {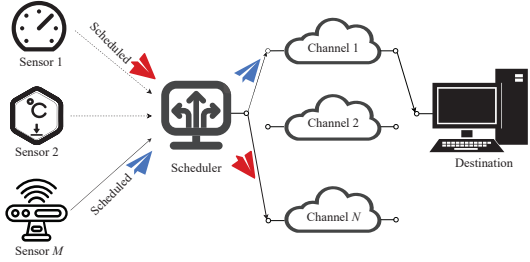}
	\caption{System Model.} 
	\label{Fig:system}
\end{figure}

We consider a discrete-time state updating network where $M$ sources send time-sensitive information to a destination through a scheduler over $N$ unreliable wireless channels, as illustrated in Fig.~\ref{Fig:system}. Let the time be slotted $t\in\{1,2,\cdots,T\}$ such that a single packet transmission occupies exactly one time slot, and $T$ is the time horizon.

In each slot, packets from each source are generated according to an independent and identically distributed (i.i.d.) Bernoulli distribution with parameter $\lambda$. Each source maintains a queue to store the packets and only keeps the latest generated packet for transmission. This discipline ensures that each updating provides the freshest information for the destination.

For each channel $n$, denote $\mu_n(t)$ to be the successful transmission probability, which is time varying due to the channel condition. We assume $\mu_n(t)$ is unknown to the scheduler in advance. However, different from \cite{agebandit2021}, we assume that the scheduler can obtain sided channel information such as SNR by pilot probing and other channel estimation methods, for example using the channel reciprocity in the Time Division Duplexing (TDD) scheme. Denote vector $\bm{b}_n(t) \in \mathbb{R}^p$ to be the gathered $p$-dimension information for channel $n$ in slot $t$, which may include SNR, interference power, etc.

We assume that the scheduler can use these observed information to construct a specific function for approximating the unknown $\mu_n(t)$ \cite{CB2021Jung,perez2010cross}. Without loss of generality, define $\mu_n(t) = \phi(\bm{b}_n(t))^T \bm{\theta} + \epsilon_t$ to be the fitting function, where $\phi:\mathbb{R}^p \to \mathbb{R}^d$ is the pre-processing function, $\bm{\theta}\in \mathbb{R}^d$ is the linear coefficient and $\epsilon_t$ is the zero-mean sub-Gaussian noise.
\begin{rem}
    Despite the linear formulation displayed above, this expression can represent multiple non-linear mappings due to the arbitrariness of $\phi$. In Section \ref{sub-opt}, we will discuss how to find the function $\phi$. Moreover, the approximation noise is restricted to be unbiased (zero-mean) in this model. This assumption is mainly for the theoretical analysis, and is not necessary for our proposed algorithm.
\end{rem}

Due to the interference constraint, we assume that one channel can only serve at most one sensor at a time. Therefore, the scheduler can select at most $p$ ($p \le \min\{M,N\}$) disjoint pairs $(m(t),n(t))$ for transmission in each slot $t$, where $m(t)\in\{1,2,\cdots,M\}$ and $n(t)\in \{1,2,\cdots,N\}$. Here for representation, in the following we consider the scheduler can only select at most one pair, i.e., $p=1$. This model is concise for analysis and expression but remains illuminating and representative because we can aggregate the potential sensors and channels subsets to be scheduled as a super-sensor and a super-channel. Then the general model is equivalent to this single-pair model. For example, in Fig. \ref{Fig:system}, if $p=2$, then we can construct the super-sensor set as $\{\{1\},\{2\},\{3\},\{1,2\},\{1,3\},\{2,3\}\}$, where \{1,2,3\} is the index of the three sensors. The construction of the super-channel is similar.

\subsection{Age of Information}
We apply the AoI metric to measure the information freshness at the destination. By definition, the AoI is the difference between the current time and the time slot when the newest packet at the destination is generated. Denote $x_m(t)$ to be the AoI of source $m$, and $g_m(t)$ to be the generation time of the latest packet from source $m$ received by the destination. Then we have $x_m(t):=t-g_m(t)$.

For measuring the information timeliness of the entire network, we consider the expected AoI of all sources, i.e.,
\begin{equation}\label{Eq:x_bar}
	\bar{x}(T) = \frac{1}{MT} \mathbb{E}\left[\sum_{t=1}^T \sum_{m=1}^M x_m(t) \right],
\end{equation}
where the expectation is taken over the randomness of channel state and scheduling decision of $(m,n)$.

\subsection{Optimization Problem}\label{sub-opt}
The scheduler schedules pairs $(m,n)$ in order to minimize the expected AoI $\bar{x}(T)$ for the entire network. This includes the source policy and the channel policy, denoted by $\pi_s$ and $\pi^c$ respectively. In this paper, we consider the joint scheduling policy to be a composition of a source policy and a channel policy, denoted by $\pi_s^c$ (or just $\pi$ for simplicity).

For the source policy, previous works \cite{TMC2021Eytan} have studied the age minimization problem in the multi-source network with known statistics and proposed some low-complexity and near-optimal source policies. Therefore, in the following, we analyze the performance of the optimal source policy and different channel policies. Specifically, the optimal source policy, denoted by $\pi_{s^\star}$, is defined as follows:
\begin{equation}
	\pi_{s^\star} = \arg \min_{\pi_s} \mathbb{E}\left[\sum_{t=1}^T \sum_{m=1}^M x_m^{(\pi_s^c)}(t) \right],~\forall \pi^c,
\end{equation} 
where $x_m^{(\pi_s^c)}(t)$ is the AoI of source $m$ under the scheduling policy $\pi_s^c$.

For the channel policy, the scheduler should identify the optimal channel $n^\star(t)$ with maximum successful transmission probability, denoted by $\mu^\star(t)$. To meet the requirement, we resort to the online sequential decision making framework, and formulate the scheduling strategy into a linear contextual bandit (LCB) problem. However, there are still some difficulties which prevent the standard LCB techniques from being applied to the channel selection problem.

First, we should carefully construct the pre-processing function $\phi(\cdot)$. In the standard LCB problem, the algorithm estimates the parameter $\bm{\theta}$ with the knowledge of $\phi$. This requires the scheduler to find the potential features which are linear with $\mu_n(t)$. In practice, the scheduler may construct $\phi$ through prior information, or list multiple possible linear features for approximation (like Taylor expansion). In the following, we will assume the linear regression is unbiased and in the simulation we will consider both cases and study the influence of the biased error.

Next, we introduce the new AoI regret metric. In online learning algorithms, regret is defined as the reward difference between the proposed policy and the optimal policy in hindsight \cite{OCO2012Shai}. An important feature of the reward structure in the standard LCB problem is i.i.d. over time and only depends on the current action and observation. However, in our problem, the AoI metric not only depends on the current action pair $(m,n)$, but also depends on the historical decisions and results. Therefore, similar to \cite{AoIbandit2020,agebandit2021}, we introduce the AoI regret to evaluate any scheduling policy $\pi$, denoted by $R(T)$:
\begin{equation}
	R(T) = \mathbb{E}\left[\sum_{t=1}^T \sum_{m=1}^M x_m^{(\pi)}(t) -\sum_{t=1}^T \sum_{m=1}^M x_m^{(\pi^\star)}(t)\right],
\end{equation}
where $x_m^{(\pi^\star)}(t)$ is the AoI under the optimal scheduling policy $\pi^\star$. Then, minimizing Eq.~\eqref{Eq:x_bar} is equivalent to minimizing the AoI regret.

\begin{rem}
    Although the AoI regret definition is similar to \cite{AoIbandit2020,agebandit2021}, the optimal policy $\pi^\star$ chosen for comparison is different. In our work, $\pi^\star$ is to choose the optimal source and channel pair in each slot, while in the previous work $\pi^\star$ is to choose one fixed optimal pair in the entire horizon. Obviously, the benchmark used in our work is more superior, which indicates that considering the sided information can greatly improve the AoI performance.
\end{rem}

\section{Problem Resolution}
\subsection{AoI Regret Bound}
Before proposing the algorithms, first we study the AoI regret and find the relationship between the standard LCB regret and the history-dependent AoI regret. Notice that under the same optimal source policy, the AoI regret results from choosing the sub-optimal channel for transmission. Denote $\mathbb{E}\left[K^\pi(T)\right]$ to be the expected times of sub-optimal channel choices over the time horizon under policy $\pi$. Then we have the following theorem:
\begin{theorem}\label{Thm:R(T)}
	Assume $\mu_n(t)$ is lower bounded by a positive constant, i.e., $\mu_n(t) \ge \mu_{\text{min}},~\forall n,t$. 
	
	For $p=1$, if the optimal channel is distinguishable enough, i.e., there exists a positive constant $\Delta_{\text{min}}$ such that $\mu^\star(t) -\mu_n(t)\ge \Delta_{\text{min}}$ for any sub-optimal channel $n$ in each slot, then the AoI regret under the optimal source policy $\pi_{s^\star}$ and any channel policy $\pi^c$ scales with $\mathbb{E}\left[K^\pi(T)\right]$, i.e.,
	\begin{equation}
		R(T) = \Theta(\mathbb{E}\left[K^\pi(T)\right]). \footnote{$f(n)=\Theta(g(n))$ means for f(n), if there exists positive $n_0, c_1, c_2$ such that when $n\ge n_0$, $0 \le c_1g(n) \le f(n) \le c_2g(n)$}
	\end{equation}

	For the general case $p > 1$, the AoI regret under the optimal source policy $\pi_{s^\star}$ and any channel policy $\pi^c$ grows at most with $\mathbb{E}\left[K^\pi(T)\right]$, i.e.,
	\begin{equation}
		R(T) = \mathcal{O}(\mathbb{E}\left[K^\pi(T)\right]).
		\footnote{$f(n)=\mathcal{O}(g(n))$ means for f(n), if there exists positive $n_0, c$ such that when $n\ge n_0$, $0 \le f(n) \le cg(n)$}
	\end{equation}
\end{theorem}
\begin{IEEEproof}
	Due to the space limitation, the proof is provided in the technical report. The outline is based on \cite[Proposition 2 and 3]{agebandit2021}, but needs some modification and extension to be applied to our system. The limitation of \cite{agebandit2021} includes not considering the sided information and just giving proof outline of $p=1$. In fact, introducing the sided information and proving the bound (especially the lower bound) need more delicate scaling. Here we give an intuitive analysis. If we choose the optimal source policy, then the AoI regret comes from choosing a sub-optimal channel for transmission. For each sub-optimal channel selection, it causes a bounded AoI increase. Therefore, the AoI regret scales with the total number of sub-optimal selections. However, the rigorous proof requires more delicate construction since under the same source policy, the source selection in each slot may still be different.
\end{IEEEproof}

\subsection{Age-Independent Policy}
Theorem \ref{Thm:R(T)} indicates that minimizing AoI regret is somehow equivalent to minimizing sub-optimal channel selection times. Notice that the channel selection is independent among different slots. Therefore, a natural design of channel policies can be based on traditional LCB algorithms including Linear UCB (LinUCB) and Linear TS (LinTS). We propose a jointly scheduling scheme by employing LinUCB and LinTS, as summarized in Algorithm \ref{alg:LCB}. Without loss of generality, let the pre-processing $\phi(\bm{x})=\bm{x}$, i.e., the information observed has already been processed.
\begin{algorithm}
	\caption{Scheduling algorithm based on LinUCB or LinTS channel policy}\label{alg:LCB}
	\begin{algorithmic}[1]
		\STATE \textbf{Initialization}: $\bm{A}=\bm{I}_d$ (identity matrix), $\bm{b}=\bm{0}$;
		\STATE \textbf{LinUCB} parameter: $\alpha$; \textbf{LinTS} parameter: $v$;
		\FOR{$t=1,2,\cdot,T$}
		\IF{there is a new packet for at least one source}
		\STATE \textbf{[Source policy]:} Select $m(t)\in\{1,2,\cdots,M\}$ according to the optimal source policy;
		\STATE Observe $N$ sided information: $\bm{b}_n(t)$;
		\STATE Coefficient estimation: $\bm{\theta}_t = \bm{A}^{-1} \bm{b}$;
		\STATE \textbf{[Channel policy]:}
		\STATE \textbf{LinUCB:} Select channel $n(t) = \arg \max [\bm{\theta}_t^T \bm{b}_n(t) + \alpha \sqrt{\bm{b}_n^T(t)\bm{A}^{-1}\bm{b}_n(t)}]_0^1$;
		\STATE \textbf{LinTS:} Sample $\tilde{\bm{\theta}}_t$ from $\mathcal{N}(\bm{\theta}_t,v^2 \bm{A}^{-1})$, and select channel $n(t) = \arg \max [\tilde{\bm{\theta}}_t^T \bm{b}_n(t)]_0^1$;
		\STATE Observe transmission result $r_t\in\{0,1\}$;
		\STATE Update parameter: $\bm{A}=\bm{A}+\bm{b}_{n(t)}(t)\bm{b}_{n(t)}^T(t)$; $\bm{b} = \bm{b} + \bm{b}_{n(t)}(t)r_t$; 
		\ENDIF
		\ENDFOR
	\end{algorithmic}
\end{algorithm}

Different from standard LinUCB and LinTS, in step 9 and 12 in Algorithm \ref{alg:LCB}, the estimated value is projected into the interval $[0,1]$. This is because the successful probability is bounded in $[0,1]$. The parameters of $\alpha$ (LinUCB) and $v$ (LinTS) are determined to guarantee a high probability regret bound. As proven in \cite{pmlr-chu11a} and \cite{pmlr-agrawal13}, $\alpha$ is chosen as $\sqrt{\frac{1}{2} \ln \frac{2TN}{\delta}}$ to achieve $\mathcal{O}(d\sqrt{T \ln((1+T)/\delta)})$ bound and $v$ is chosen as $\sqrt{\frac{24}{\epsilon}d\ln\frac{1}{\delta}}$ to achieve $\mathcal{O}( \frac{d^2}{\epsilon}\sqrt{T^{1+\epsilon}}\ln(Td)\ln\frac{1}{\delta})$ bound with probability $1-\delta$. However, in practice, the parameter $v$ can be chosen smaller for more cautious exploration to achieve better performance.

\subsection{Age-Dependent Policy}

Notice that the standard LCB-based channel selection algorithms are unaware of the current AoI. We call these age-independent policies. This motivates us to combine the AoI information with the algorithm to achieve better performance, called age-dependent policies. Intuitively, it is more reasonable to exploit the current estimation when the AoI is high and to explore other possibilities when the current age is low. To exploit the AoI information, we propose two age-dependent channel selection strategies based on the original Algorithm \ref{alg:LCB}, as summarized in Algorithm \ref{alg:AD-LCB}.
\begin{algorithm}
	\caption{Scheduling algorithm based on age-dependent channel policy AD-UCB or AD-TS}\label{alg:AD-LCB}
	\begin{algorithmic}[1]
		\STATE \textbf{Initialization}: $\bm{A}=\bm{I}_d$ (identity matrix), $\bm{b}=\bm{0}$;
		\STATE \textbf{AD-UCB} parameter: $\alpha$; \textbf{AD-TS} parameter: $v$;
		\FOR{$t=1,2,\cdot,T$}
		\IF{There is a new packet for at least one source}
		\STATE \textbf{[Source policy]:} Select $m(t)\in\{1,2,\cdots,M\}$ according to the optimal source policy;
		\STATE Observe $N$ sided information: $\bm{b}_n(t)$;
		\STATE \textbf{[Channel policy]:}
		\IF{$x_m(t) > \frac{M}{2\lambda \max(\bm{\theta}_t^T \bm{b}_n(t))}$}
		\STATE Select channel $n(t) = \arg \max \bm{\theta}_t^T \bm{b}_n(t)$;  // \textbf{totally exploitation}
		\ELSE
		\STATE Select channel $n(t)$ as Algorithm \ref{alg:LCB} step 9 or 10; // \textbf{bandit exploitation and exploration}
		\ENDIF
		\STATE Observe transmission result $r_t\in\{0,1\}$;
		\STATE Update parameter: $\bm{A}=\bm{A}+\bm{b}_{n(t)}(t)\bm{b}_{n(t)}^T(t)$; $\bm{b} = \bm{b} + \bm{b}_{n(t)}(t)r_t$; 
		\ENDIF
		\ENDFOR
	\end{algorithmic}
\end{algorithm}

The intuition of Algorithm \ref{alg:AD-LCB} is to set an AoI threshold determining whether to follow the bandit algorithm (step 11) or to directly exploit (step 9). The threshold $\frac{M}{2\lambda \max(\bm{\theta}_t^T \bm{b}_n(t))}$ is chosen as the average AoI under the round robin policy with the maximum estimated successful probability $\max(\bm{\theta}_t^T \bm{b}_n(t))$.

\begin{rem}
	Both the proposed Algorithms can be easily generalized to the multi-pair scheduling scenario where $p > 1$ by applying the combinatorial bandit method \cite{Chen2013CMAB} while satisfying the regret bound. An intuitive idea is to consider all the possible scheduling channel subsets as super-channels, and the scheduler selects a super-channel in each slot, which degenerates to the $p=1$ model. Another low-complexity method is to pick the largest $p$ channels in the proposed algorithms, which also achieves good performance demonstrated in the simulation part.
\end{rem}

\section{Simulation Results}

In this section, we provide some simulation results to demonstrate the performance of proposed algorithms. We compare the channel algorithms by employing the Age-based Max-Weight source policy \cite{TMC2021Eytan}, which selects the source $m$ for transmission with the packet causing the largest AoI reduction under a successful transmission. It is shown in \cite{TMC2021Eytan} that this source policy can achieve near optimal AoI performance.

\begin{figure}
	\centering
	\includegraphics[width=0.8\linewidth]{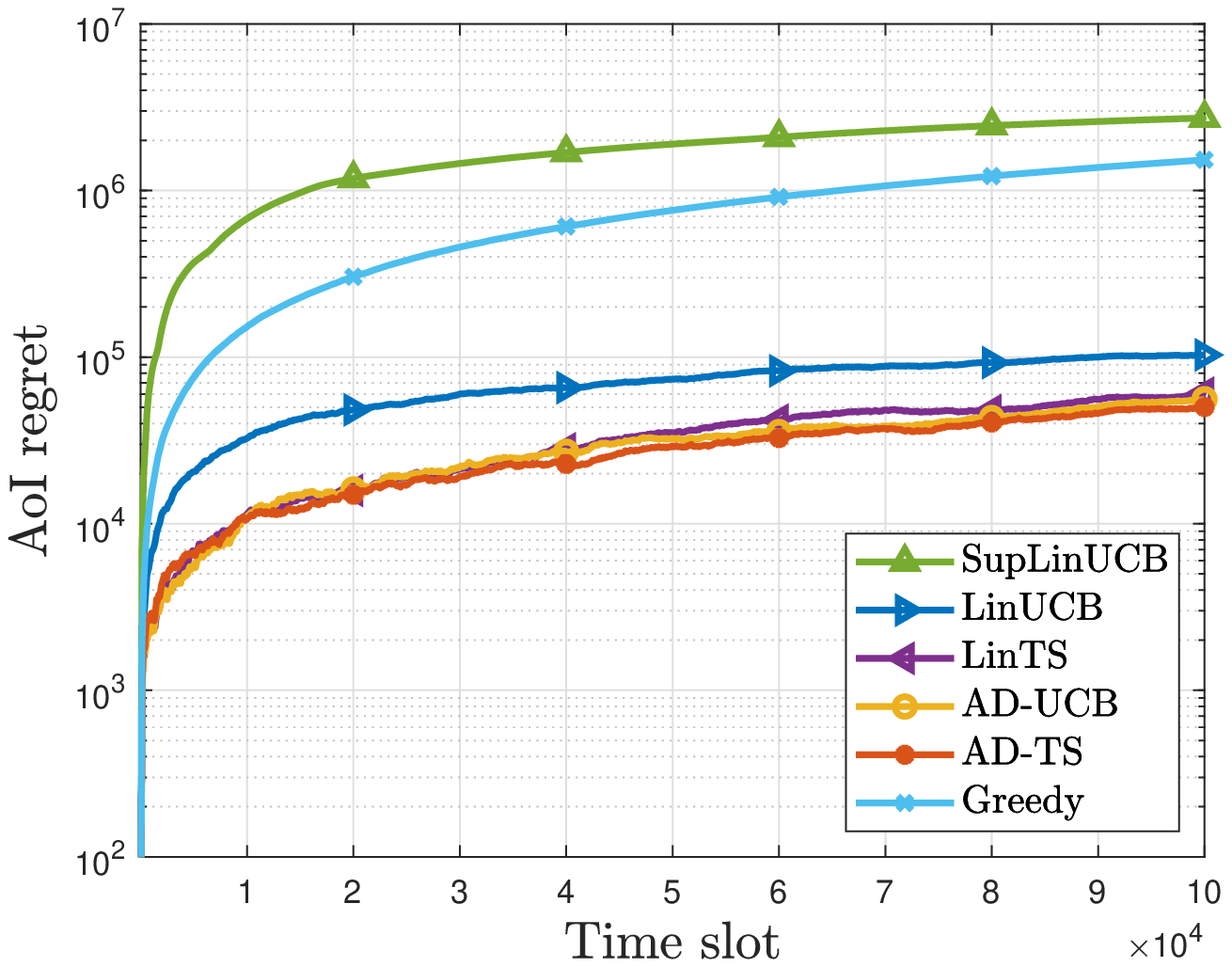}
	\caption{AoI regret performance in linear contextual case.}
    \label{Fig:linear}
\end{figure}

We simulate a network with $M=20$ sources and $N=5$ channels with time horizon $T=10^5$ slots. The performance is evaluated by running 1000 rounds for average. The packet arrival rate for each source is $\lambda=0.5$. First, we consider the linear contextual model with the true parameter $\bm{\theta} = [0.9,0.1,0.7]$ and the noise follows the uniform distribution $U(-0.03,0.03)$. The contextual information for each channel varies as summarized in Table \ref{tab:dist}, where $\delta(x)$ is the impulse function, $U(a,b)$ is the uniform distribution on interval $[a,b]$ and $T(a,b,c)$ is the triangle distribution on interval $[a,b]$ with mode $c$. The AoI regret performance is depicted in Fig.~\ref{Fig:linear} compared with the modified SupLinUCB proposed in \cite{pmlr-takemura21a}, where the parameters $v=1$ in LinTS and $\alpha=\sqrt{\frac{1}{2} \ln \frac{2TN}{\delta}}$ in LinUCB. From Fig.~\ref{Fig:linear}, we can verify the sub-linear regret growth for the AoI regret, and the improved performance of the age-dependent policies.

\begin{figure}
	\centering
	\includegraphics[width=0.8\linewidth]{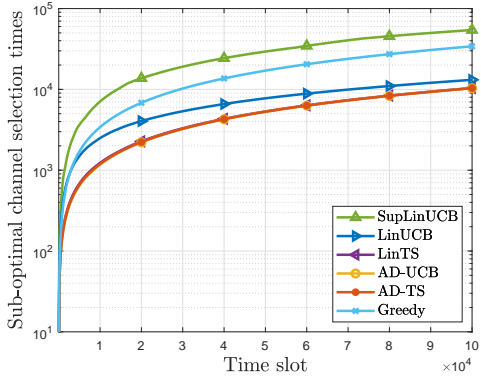}
	\caption{Sub-optimal channel selection times in linear contextual case.}
    \label{Fig:channel}
\end{figure}

To validate Theorem \ref{Thm:R(T)}, we draw Fig.~\ref{Fig:channel}. From Fig.~\ref{Fig:linear} and Fig.~\ref{Fig:channel}, we can see that the AoI performance is closely related to the sub-optimal channel selection times, as concluded in Theorem \ref{Thm:R(T)}. Moreover, the AoI regret roughly scales with the number of sub-optimal selections. This is intuitive since all the policies apply the same source selection policy, and thus the AoI regret mainly comes from sub-optimal channel selections.

\begin{table}[ht]\scriptsize
	\begin{center}
		\caption{Distribution of different channel sided information}
		\label{tab:dist}
		\begin{tabular}{llll}
			\toprule
			Channel & Dimension 1 & Dimension 2 & Dimension 3 \\
			\midrule
			1 & $\delta(x-0.4)$ & $\delta(x-0.8)$ & $\delta(x-0.2)$ \\
			2 & $U(0,0.3)$ & $U(0,2.5)$ & $U(0,0.6)$ \\
			3 & $T(0,0.3,0.15)$ & $T(0,2.4,1.2)$ & $T(0,0.6,0.3)$ \\
			4 & $0.3\delta(x-0.4)$ & $0.3\delta(x-3.5)$ & $0.3\delta(x-0.3)$ \\
			$ $ & $+0.7\delta(x-0.2)$ & $+0.7\delta(x-1.5)$ & $+0.7\delta(x-0.4)$ \\
			5 & $0.5\text{Beta}(3,4)$ & $3\text{Beta}(3,4)$ & $0.2\text{Beta}(3,4)$ \\
			\bottomrule
		\end{tabular}
	\end{center}
\end{table}

\begin{figure}
	\centering
	\includegraphics[width=0.8\linewidth]{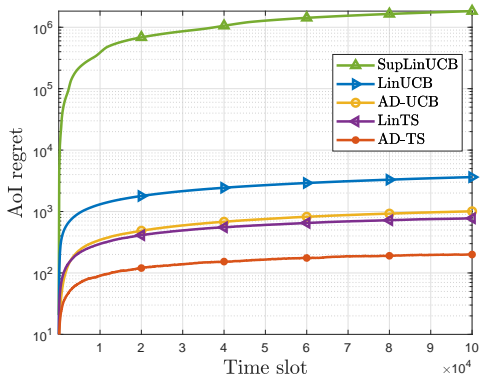}
	\caption{AoI regret performance in non-linear contextual case.}
	\label{Fig:sigmoid}
\end{figure}

Next, we consider the non-linear case depicted in Fig.~\ref{Fig:sigmoid}. Here we use the exponential approximation proposed in \cite[Chapter 2.5]{perez2010cross}, i.e., the successful probability $\mu_n=1 - \exp(-(\gamma+2))$, where $\gamma$ is the SNR (dB) information following a uniform distribution $U(-2,6)$. When the true model becomes non-linear, it is impossible to achieve sub-linear AoI regret by the LCB-based algorithm due to the modeling error. However, the proposed algorithms still beat the modified SupLinUCB proposed under the misspecified LCB setting, and the age-dependent policies can still improve the performance by forcing exploitation in the high AoI time slots.

\begin{figure}
	\centering
	\includegraphics[width=0.8\linewidth]{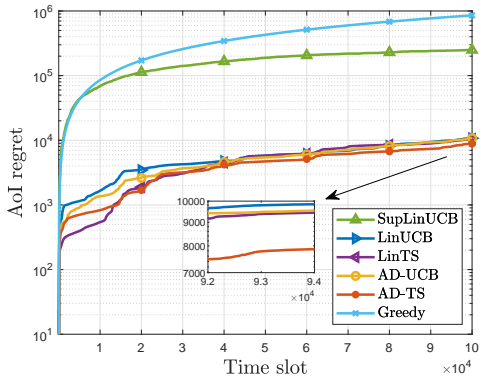}
	\caption{AoI regret performance with $p=3$.}
	\label{Fig:multipair}
\end{figure}

Finally, we consider the multi-pair scheduling case, and choose the linear model as Fig. \ref{Fig:linear} and $p=3$. Here we use the low-complexity generalized algorithms mentioned in Remark 3. As depicted in Fig. \ref{Fig:multipair}, our proposed algorithms still outperform the SupLinUCB algorithm, but the difference between the four algorithms becomes small. This may be because the scheduling freedom is larger when $p=3$, which can greatly reduce the AoI compared with $p=1$.

\section{Conclusion}
We consider an age optimization problem in a multi-source multi-channel network, where only sided channel state information can be observed. Different channel learning algorithms are proposed based on the linear contextual bandit to achieve sub-linear AoI regret compared with the optimal policy. We also demonstrate the performance of the proposed algorithms under the non-linear case, and find the benefit of considering the current AoI for channel decision. Interesting extensions include the time-varying channel model parameters $\bm{\theta}$, and the design of the more general scheduling policies where source and channel selection decisions may be coupled.

\appendices
\section{Proof of Theorem \ref{Thm:R(T)} when $p=1$}\label{Proof:Thm1}
Similar to \cite{queuebandit2021}, we introduce the coupled channels to complete our proof.
\begin{Def}[Coupled Channels]
	Denote $c_n(t)$ to be the indicate variable, where $c_n(t)=1$ means the successful transmission through channel $n$ in slot $t$ and $c_n(t)=0$ otherwise. For simplicity, denote $c^\star(t)$ to be the optimal channel indicate variable. Let $\{U(t)\}_{t=1}^T$ to be a sequence of i.i.d. uniformly distributed random variables in $[0,1]$. Then the coupled channels means $c_n(t)=1$ in each slot $t$ if and only if $0\le U(t) \le \mu_n(t),~\forall n$.
\end{Def}

By definition, the coupled channels indicate that if a transmission over channel with a low $\mu_n(t)$ is successful, then so is the transmission over channels with a higher $\mu_n(t)$. Meanwhile, the marginal distribution of the successful transmission is the same as the original channel. Therefore, the distribution of AoI also remains the same for both coupled and original channels. This coupled property will help us analyze the AoI regret in the following.

First, we prove the lower bound, i.e., $R(T) = \Omega(\mathbb{E}\left[K^\pi(T)\right])$. Notice that although both $\pi$ and the optimal policy use the same source policy, the source choice in each slot may not be the same. To overcome this challenge, we construct another policy $\hat{\pi}$, which selects the same source as $\pi$ but choose the optimal channel. Then, the AoI regret $R(T)$ can be lower bounded by:
\begin{equation}\label{Eq:R(T)_lower}
	R(T) \ge \mathbb{E}\left[\sum_{t=1}^T \sum_{m=1}^M x_m^{(\pi)}(t) -\sum_{t=1}^T \sum_{m=1}^M x_m^{(\hat{\pi})}(t)\right].
\end{equation}

Let $w_m^{(\pi)}(t)$ be the indicator variable whether the newest packet of source $m$ is received by the destination in slot $t$ under policy $\pi$. Then the AoI evolution can be written as:
\begin{align*}
	x_m^{(\pi)}(t) = (1-w_m^{(\pi)}(t))(x_m^{(\pi)}(t-1)+1)+w_m^{(\pi)}(t)\tau_m(t), \\
	x_m^{(\hat{\pi})}(t) = (1-w_m^{(\hat{\pi})}(t))(x_m^{(\hat{\pi})}(t-1)+1)+w_m^{(\hat{\pi})}(t) \tau_m(t),
\end{align*}
where $\tau_m(t)$ is the time elapsed since the generation of the newest packet of source $m$, and is independent of the scheduling policy. Notice that $\tau_m(t)$ is not equivalent to $x_m(t)$ because the newest packet may not be received by the destination. In the coupled system, we always have $w_m^{(\pi)}(t) \le w_m^{(\hat{\pi})}(t)$. Therefore, $x_m^{(\pi)}(t) \ge x_m^{(\hat{\pi})}(t)$, and the difference can be bounded as
\begin{align*}
	&x_m^{(\pi)}(t) - x_m^{(\hat{\pi})}(t) \\
	=& (1-w_m^{(\pi)}(t))x_m^{(\pi)}(t-1) - (1-w_m^{(\hat{\pi})}(t))x_m^{(\hat{\pi})}(t-1) \\
	&+(1-\tau_m(t))(w_m^{(\hat{\pi})}(t)-w_m^{(\pi)}(t)) \\
	\ge& (w_m^{(\hat{\pi})}(t)-w_m^{(\pi)}(t))(x_m^{(\hat{\pi})}(t-1)+1-\tau_m(t)).
\end{align*}

Taking expectation on both sides yields
\begin{align}\label{Eq:x_and_s}
	&\mathbb{E}\left[x_m^{(\pi)}(t) - x_m^{(\hat{\pi})}(t) \right] \nonumber\\
	\overset{(a)}{\ge}& \mathbb{E}\left[w_m^{(\hat{\pi})}(t)-w_m^{(\pi)}(t) \right]  \mathbb{E}\left[x_m^{(\hat{\pi})}(t-1)+1-\tau_m(t) \right] \nonumber\\
	\overset{(b)}{\ge}& \mathbb{E}\left[w_m^{(\hat{\pi})}(t)-w_m^{(\pi)}(t) \right],
\end{align}
where (a) holds since the $x_m^{(\hat{\pi})}(t-1)-\tau_m(t)$ and $w_m^{(\hat{\pi})}(t)-w_m^{(\pi)}(t)$ are independent, and (b) holds since $x_m^{(\hat{\pi})}(t-1)+1$ is always larger than $\tau_m(t)$. Notice that $w_m^{(\hat{\pi})}(t)-w_m^{(\pi)}(t)=1$ results from a sub-optimal channel choice in the current slot $t$ or a previous slot, denoted by $\tau(t)$. Conversely, in some slot $\tau$, the event $m(\tau)=m$ and $\mu_{n(\tau)}(\tau)<U(\tau)\le\mu^\star(\tau)$ will cause desynchronization of $w_m^{(\hat{\pi})}(\tau')$ and $w_m^{(\pi)}(\tau')$ for several slots $\tau'>\tau$ in the coupled system. Then we can lower bound $w_m^{(\hat{\pi})}(t)-w_m^{(\pi)}(t)$ as
\begin{align*}
	&\mathbb{E}\left[w_m^{(\hat{\pi})}(t)-w_m^{(\pi)}(t) \right] \\
	\ge& \mathbb{E}\left[\sum_{n=1}^N \mathbb{I}\{m(\tau(t))=m,n(\tau(t))=n\}(c^\star(\tau(t))-c_n(\tau(t))) \right]\\
	=& \sum_{n\neq n^\star(\tau(t))} \mathbb{P}\left(\mathbb{I}\{m(\tau(t))=m,n(\tau(t))=n\} \right)\\ 
	&\cdot \mathbb{P}\left( \mu_{n}(\tau(t))<U(\tau(t))\le\mu^\star(\tau(t))\right) \\
	\ge& \Delta_n(\tau(t)) \cdot \sum_{n\neq n^\star(\tau(t))} \mathbb{P}\left(\mathbb{I}\{m(\tau(t))=m,n(\tau(t))=n\} \right),
\end{align*}
where $\mathbb{I}(\cdot)$ is the indicator function, and $\Delta_n(\tau(t)) = \min_{n\neq n^\star(\tau(t))} (\mu^\star(\tau(t)) - \mu_n(\tau(t)))$. Summing up all the sources, we have:
\begin{align}
	\sum_{m=1}^M \mathbb{E}\left[w_m^{(\hat{\pi})}(t)-w_m^{(\pi)}(t) \right] \ge \Delta_{\text{min}} \mathbb{E}\left[\mathbb{I}\{n(\tau(t))\neq n^\star\} \right].
\end{align}

Summing up $t$ and recalling Eq.~\eqref{Eq:R(T)_lower} and Eq.~\eqref{Eq:x_and_s} yield $R(T)=\Omega(\mathbb{E}\left[K^\pi(T)\right])$.

Next, we prove the upper bound, i.e., $R(T) = \mathcal{O}(\mathbb{E}\left[K^\pi(T)\right])$. Similar to the lower bound proof, we construct an auxiliary policy $\tilde{\pi}$ which selects the source as $\pi^\star$ but selects the channel as $\pi$. Since $\pi$ applies the optimal source policy, the expected AoI under $\pi$ is no greater than $\tilde{\pi}$. Then the regret can be upper bounded as
\begin{equation}\label{Eq:R(T)_upper}
	R(T) \le \mathbb{E}\left[\sum_{t=1}^T \sum_{m=1}^M x_m^{(\tilde{\pi})}(t) -\sum_{t=1}^T \sum_{m=1}^M x_m^{(\pi^\star)}(t)\right].
\end{equation}

Notice that the AoI regret between $\tilde{\pi}$ and $\pi^\star$ results from a sub-optimal choice and $\mu_{n(t)}(t)<U(t)\le \mu^\star(t)$. Suppose the choice happens when transmitting packet from source $m$ in slot $t$. Then after that, there causes an AoI discrepancy until at the next time when the packet from source $m$ is successfully transmitted under $\tilde{\pi}$. Denote the length of the discrepancy to be $L_t$. Then the AoI regret can bounded as
\begin{align}\label{Eq:Upper_C}
	R(T)\le &\sum_{t=1}^T \sum_{m=1}^M \mathbb{E}\left[\mathbb{I}\{n(t)\neq n^\star,m(t)=m \}\right]\nonumber \\
	 &\mathbb{E}\left[\frac{1}{2}L_t^2 + L_t x_m^{(\tilde{\pi})}(t-1) \right] \mathbb{P}(\mu_{n}(t)<U(t)\le \mu^\star(t)),
\end{align}
where the second term of RHS is the cumulative AoI under $\tilde{\pi}$ during the discrepancy period. Notice that both $\tilde{\pi}$ and $\pi^\star$ select the same source in each slot, and any feasible policy will schedule each source $m$ within a finite slots, bounded by a constant $C$. Otherwise, the AoI of source $m$ will be infinite, which is far from optimality. For each time slot of scheduling source $m$, the number of times the packet is transmitted until success is dominated by another random variable $C\cdot Y$, where $Y$ is geometrically distributed with parameter $\frac{1}{\mu_{\text{min}}}$. Then we have $\mathbb{E}[L_t] \le \frac{C}{\mu_{\text{min}}}$, and $\mathbb{E}[L_t^2] \le C^2 \frac{2-\mu_{\text{min}}}{\mu_{\text{min}}^2}$.

Moreover, although $L_t$ and $x_m^{(\tilde{\pi})}(t-1)$ are dependent, $Y$ and $x_m^{(\tilde{\pi})}(t-1)$ are independent. Therefore, we can upper bound $\mathbb{E}\left[L_t x_m^{(\tilde{\pi})}(t-1)\right]$ as
\begin{align*}
	\mathbb{E}\left[L_t x_m^{(\tilde{\pi})}(t-1)\right] &\le C\mathbb{E} \left[Yx_m^{(\tilde{\pi})}(t-1) \right] \\
	&\le \frac{C}{\mu_{\text{min}}}\mathbb{E}\left[x_m^{(\tilde{\pi})}(t-1) \right].
\end{align*}

For $\mathbb{E}\left[x_m^{(\tilde{\pi})}(t-1) \right]$, we can decompose it into the expected AoI caused by source packet arrival interval and waiting time for transmission, both of which are finite in expectation under optimal source policy. Therefore, the second term of RHS in Eq.~\eqref{Eq:Upper_C} is bounded by a finite constant, denoted by $\tilde{C}$, which yields
\begin{equation}
	R(T)\le \tilde{C}\sum_{t=1}^T \mathbb{E}\left[\mathbb{I}\{n(t)\neq n^\star\}\right] = \mathcal{O}(\mathbb{E}\left[K^\pi(T)\right]).
\end{equation}

\section{Proof of Theorem \ref{Thm:R(T)} when $p>1$}
For the general case, Eq.~\eqref{Eq:R(T)_upper} also holds. Denote $n_m^\star(t)$ to be the optimal channel choice for source $m$ in slot $t$, and the definition of $\mu_m^\star(t)$ is similar. Notice that $n_m^\star(t)$ may not be $n^\star(t)$ since $n^\star(t)$ can be occupied by other source $m'$. Then the AoI discrepancy results from two cases:

\textbf{Case 1 (Real sub-optimal choice):} This means in some slot $t$, the policy $\tilde{\pi}$ choose the channel such that $\mu_{n_m(t)}(t)<U(t)\le \mu_m^\star(t)$. In this case, the policy $\tilde{\pi}$ will cause a larger AoI , which is called real sub-optimal.

\textbf{Case 2 (Fake sub-optimal choice):} This means the policy $\tilde{\pi}$ choose the channel such that $\mu_m^\star(t)<U(t)\le\mu_{n_m(t)}(t)$. This may happen when for example, the source $m$ with a smaller AoI is served by a better channel. Then the AoI of source $m$ will be small but the total AoI will be large. Therefore, it is still a sub-optimal choice, but we call it fake sub-optimal to distinguish from Case 1.

Since Case 2 choice will cause AoI reduction for the specific souree, we ignore this reduction and focus on the AoI growth in Case 1. Then the AoI regret can be upper bounded by:
\begin{align}\label{Eq:Upper_C_p>1}
	R(T)\le &\sum_{t=1}^T \sum_{m=1}^M \mathbb{E}\left[\frac{1}{2}L_t^2 + L_t x_m^{(\tilde{\pi})}(t-1) \right]\nonumber \\
	 &\mathbb{E}\left[\mathbb{I}\{n_m(t)\neq n_m^\star,m(t)=m,\mu_{n_m}(t)< \mu_m^\star(t) \}\right] \nonumber \\
	 &\mathbb{P}(\mu_{n_m}(t)<U(t)\le \mu_m^\star(t)),
\end{align}

Then similar to Appendix A, $\mathbb{E}\left[\frac{1}{2}L_t^2 + L_t x_m^{(\tilde{\pi})}(t-1) \right]$ can be upper bounded by a constant $\tilde{C}$, which yields
\begin{align}
    R(T)&\overset{(a)}{\le}\tilde{C}\sum_{t=1}^T \sum_{m=1}^M \mathbb{E}\left[\mathbb{I}\{n_m(t)\neq n_m^\star,m(t)=m \}\right] \nonumber \\
    &= \mathcal{O}(\mathbb{E}\left[K^\pi(T)\right]),
\end{align}
where (a) holds since we upper bound $\mathbb{P}(\mu_{n_m}(t)<U(t)\le \mu_m^\star(t))$ as 1 and do not distinguish Case 1 ($\mu_{n_m}(t)< \mu_m^\star(t)$).

\emph{Remark:} Compared with Appendix A, we do not provide the lower bound proof for general $p>1$ case. This is mainly because the existence of case 2 fake sub-optimal choice. The intuition is that the AoI growth caused by case 1 can be lower bounded by a sub-optimal choice times a constant, but the other fake sub-optimal choice causes the AoI decrease. Then the total AoI regret cannot be lower bounded by $\mathbb{E}\left[K^\pi(T)\right]$.

\bibliographystyle{IEEEtran}
\bibliography{bibfile}

\end{document}